\documentclass[showpacs,prd,twocolumn,amsmath,amssymb]{revtex4-1}
\usepackage{graphicx}
\usepackage{amsmath,amsfonts}
\usepackage{dcolumn}
\usepackage{bm}
\usepackage{slashed}
\usepackage{nicefrac}
\usepackage{epstopdf}
\usepackage{color}
\usepackage{tikz-feynman,contour}
\def\bra{\langle}
\def\ket{\rangle}

\def\Tr{\mathrm{Tr}}
\def\cD{\mathcal{D}}
\def\cN{\mathcal{N}}

\def\cA{\mathcal{A}}
\def\cH{\mathcal{H}}

\def\R{\mathbb{R}}
\def\Ei{\mathrm{Ei}}
\def\dk#1#2{\frac{ d^{#2}{#1} }{ (2\pi)^{#2} }} 
\begin{document}
\title{Wavelet regularization of gauge theories}
\author{M. V. Altaisky}
\affiliation{Space Research Institute RAS, Profsoyuznaya 84/32, Moscow, 117997, Russia}
\email{altaisky@rssi.ru}

\date{May 6, 2020} 
\begin{abstract}
Extending the principle of local gauge invariance $\psi(x)\to \exp\left(\imath \sum_A \omega^A(x)T^A \right) \psi(x), x \in \R^d$, with $T^A$ being the generators of the gauge group $\cA$, to the fields $\psi(g)\equiv \bra \chi|\Omega^*(g)|\psi\ket$, defined on a locally compact Lie group $G$, $g\in G$, where $\Omega(g)$ is suitable square-integrable representation of $G$, it is shown that taking the coordinates ($g$) on the affine group, we get a gauge theory that is finite by construction. The renormalization group 
in the constructed theory relates to each other the charges measured at different scales. The case of the $\cA=SU(N)$ gauge group is considered.
\end{abstract}
\pacs{03.70.+k, 11.10.Hi}
                            
\keywords{Quantum field theory, renormalization, wavelets}

\maketitle

\section{Introduction \label{intro:sec}}
Gauge theories form the basis of modern high-energy physics. Quantum electrodynamics (QED) --  a quantum field theory model based on the invariance of the Lagrangian under local phase transformations of the matter fields $\psi(x) \to e^{\imath w(x)} \psi(x)$ -- was the first theory to succceed in describing the effect of the vacuum energy fluctuations on atomic phenomena, such as the Lamb shift, with an extremely high accuracy of several decimal digits \cite{Dyson2007}. The crux of  QED is that in representing the matter fields by square-integrable functions in Minkowski space it yields formally infinite Green functions, unless a special procedure, called {\em renormalization}, is applied to the action functional \cite{SP1953,Bsh1956}. Much later, it was discovered that all other known interactions of elementary particles, {\sl viz.} weak interaction and strong interaction, are also described by gauge theories.
The  difference from QED consists in the fact that the {\em multiplets} of matter fields   are transformed by unitary  matrices $\psi(x) \to U(x)\psi(x)$, making the theory non-Abelian. Due to 't Hooft, we know such theories to be {\em renormalizable}, and thus physically meaningful \cite{HooftVeltman1972}. Now they form the standard model (SM) of elementary particles -- an $\cA=SU(2) \times U(1) \times SU_c(3)$ gauge theory supplied with the Higgs mechanism of spontaneous symmetry breaking.

A glimpse  at the stream of theoretical papers in high-energy physics, from Ref. \cite{SP1953} till the present time, shows that renormalization takes a bulk of technical work, although the role of it is subjunctive to the main physical principle of gauge invariance, explicitly manifested in the existence of gauge bosons -- the carriers of gauge interaction. The role of the renormalization group (RG) is  to view the physics changing with scale in an invariant way depending on charges and parameters related to the given scale, absorbing all divergences in renormalization factors.

According to the author's point of view \cite{Altaisky2010PRD}, the cause of divergences in quantum field theory is an inadequate choice of the functional space $\mathrm{L}^2(\R^d)$. Due to the Heisenberg uncertainty principle, nothing can be measured at a sharp point: it would require an infinite momentum $\Delta p$ to keep $\Delta x \to 0$ with $\Delta p \Delta x \ge \frac{\hbar}{2}$. Instead, the values of physical fields are meaningful on a finite domain of size $\Delta x$, and hence the physical fields should be described by {\em scale-dependent functions} $\psi_{\Delta x}(x)$. As it was shown in previous papers \cite{Alt2002G24J,Altaisky2010PRD,AK2013}, having defined the fields $\psi_{a}(x)$ as the 
{\em wavelet transform}
 of square-integrable fields, we yield a quantum field theory of scale-dependent fields -- a theory finite by construction with no renormalization required to get rid of divergences. 

The present paper makes an endeavour to construct a gauge theory based on local unitary transformations of the {\em scale-dependent fields}: $\psi_a(x) \to U_a(x) \psi_a(x)$. The physical fields in such a theory are defined on a region of finite-sized $\Delta x$ centred  at $x$ as a sum of all scale components from $\Delta x$ to infinity by means of the inverse wavelet transform. The Green functions are finite by construction. The RG symmetry represents the relations between the charges measured at different scales.

This is essentially important for quantum chromodynamics, the theory of strong interactions, where the ultimate way of analyzing the hadronic interactions at both short and  long distances remains the study of the dependence 
of the coupling constant $\alpha_S$ on {\em only one} parameter -- the squared transferred momentum $Q^2$. Naturally, 
one can suggest that two parameters may be better than one. As  has been realized in classical physics of 
strongly coupled nonlinear systems -- first in geophysics \cite{GGM1984} -- the use of two parameters (scale and frequency) may solve a problem that appears hopeless for spectral analysis. Attempts of a similar kind have also been made in QCD 
\cite{Federbush1995}, although they have not received further development. 

I must admit in this respect, that one of the challenges of modern QCD is to separate the short-distance behaviour 
of quarks, where the perturbative calculations are feasible, from the long-distance dynamics of quark confinement --  
and then somehow to relate the parameter $\Lambda$, which describes the short-range dynamics, to the mass scale of 
hadrons \cite{DEUR2016}. This may not be the ultimate solution of the problem: sometimes it is easier to scan the 
whole range of scales with some continuous parameter  than to separate the small-- and the large-scale modes \cite{SD2000}.     

The remainder of this paper is organized as follows: In {\em Sec. \ref{lgt:sec}}, I briefly review the formalism 
of local gauge theories, as it is applied to the Standard Model and QCD. {\em Section \ref{sqft:sec}} summarizes the wavelet 
approach to quantum field theory, developed by the author in previous papers \cite{Alt2002G24J,Altaisky2010PRD,AK2013}, which yields finite Green functions $\bra \phi_{a_1}(x_1)\ldots\phi_{a_n}(x_n)\ket$ for scale-dependent fields. Its application to gauge theories, however, remains cumbersome.
{\em Section \ref{gis:sec}} is the main part of the paper. It presents the formulation of gauge invariance in scale-dependent formalism, set up the Feynman diagram technique, and gives the one-loop contribution in a pure gauge theory to the 
three-gluon vertex in scale-dependent Yang-Mills theory. 
The developed formalism aims to catch the effect of asymptotic freedom in a non-Abelian gauge theory that is finite by construction, and hopefully, with fermions being included, to describe the color confinement and enable analytical 
calculations in QCD. The problems and prospectives of the developed methods are summarized in the {\em Conclusion}. 

\section{Local gauge theories \label{lgt:sec}}
The theory of gauge fields stems from the invariance of the  action functional under the local phase transformations of the matter fields. Historically, it  originated in quantum electrodynamics, where the matter fields $\psi$, spin-$\frac{1}{2}$ fermions with mass $m$, 
are described by the action functional  
\begin{equation}
S_E = \int d^4x \Bigl[ \bar\psi \gamma_\mu \partial_\mu \psi 
+ \imath  m \bar\psi \psi \Bigr], \label{de}
\end{equation}
written in Euclidean notation, with  $\gamma$ matrices satisfying the anticommutation 
relations $\{\gamma_\mu,\gamma_\nu\}=-2\delta_{\mu\nu}$.

The action functional [Eq.\eqref{de}] can be made invariant under the local phase 
transformations 
\begin{equation}
\psi(x) \to U(x)\psi(x),\quad U(x) \equiv e^{\imath w(x)} \label{gt}
\end{equation}
by changing the partial derivative $\partial_\mu$ into the covariant 
derivative 
\begin{equation}
 D_\mu = \partial_\mu  + \imath A_\mu(x).
 \label{cd}
\end{equation}
The modified action 
\begin{equation}
S_E' = \int d^4x \Bigl[ \bar\psi \gamma_\mu D_\mu \psi 
+ \imath  m \bar\psi \psi \Bigr] \label{de1}
\end{equation}
remains invariant under the phase transformations of Eq.\eqref{gt} if 
the {\em gauge field} $A_\mu(x)$ is transformed accordingly:
\begin{equation}
A_\mu^U = U(x) A_\mu(x) U^\dagger(x) + \imath \left(\partial_\mu U(x) \right) U^\dagger (x). \label{at}
\end{equation}

Equation \eqref{gt} represents gauge rotations of the matter-field multiplets.  
The matrices $w(x)$ can be expressed in the basis of appropriate 
generators $$w(x)=\sum_A w^A(x) T^A,$$ where $T^A$ are the generators of the gauge group $\cA$, acting on matter fields in the fundamental 
representation. For the Lie group they satisfy the commutation relations 
$$
[T^A,T^B]=\imath f^{ABC}T^C,
$$
and are normalized as $\mathrm{Tr}[T^A,T^B]=T_F \delta^{AB}$; where $T_F=\frac{1}{2}$ is a common choice.
For the Yang-Mills theory I assume the symmetry group to be $SU(N)$. The trivial case of $N=1$ corresponds to the Abelian theory -- quantum electrodynamics.

The Yang-Mills action, which describes the action of the gauge field $A_\mu(x)$ itself, should be added to the action in Eq.\eqref{de1}. It is expressed in terms of the field-strength tensor 
\begin{equation}
S_{YM}[A] = \frac{1}{2g^2}\int \mathrm{Tr}( F_{\mu\nu}  F_{\mu\nu}) d^4x,
\end{equation}
where 
\begin{equation}
F_{\mu\nu}=-\imath [D_\mu,D_\nu] 
= \partial_\mu A_\nu - \partial_\nu A_\mu 
+ \imath [A_\mu, A_\nu] \label{fmn} 
\end{equation} 
is the strength tensor of the gauge field, and $g$ is a formal coupling constant obtained by redefinition of the 
gauge fields $A_\mu\to g A_\mu$. 
 
It should be noted that the free-field action [Eq.\eqref{de}] -- 
that has given rise to gauge theory -- was written in a Hilbert space of square-integrable functions $\mathrm{L}^2(\R^d)$, 
with the scalar product 
$
\bra \psi | \phi\ket = \int \bar{\psi}(x)\phi(x) d^d x.
$
In what follows, the same will be done for more general Hilbert spaces.

\section{Scale-dependent quantum field theory \label{sqft:sec}}
\subsection{The observation scale}
The dependence of physical interactions on the scale of observation is of paramount importance. In classical physics, 
when the position and the momentum can be measured simultaneously, one can average the measured quantities over 
a region of a given size $\Delta x$ centred at point $x$. For instance, the Eulerian velocity of a fluid, measured 
at point $x$ within a cubic volume of size $\Delta x$, is given by 
$$
v_{\Delta x}(x) := \frac{1}{(\Delta x)^d} \int_{(\Delta x)^d}v(x) d^dx.
$$

In quantum physics it is impossible to measure any field $\phi$ sharply at a point $x$. This would require an infinite momentum transfer $\Delta p \sim \hbar/\Delta x$, with $\Delta x\to0$ making $\phi(x)$ meaningless. That is why, any such field should be designated by 
the resolution of observation: $\phi_{\Delta x}(x)$. In high-energy physics experiments, the initial and final states 
of particles are usually determined in momentum basis $|p\ket$, -- the plane wave basis. For this reason, the 
results of measurements -- i.e., the correlations between different events -- are considered as functions of squared 
momentum transfer $Q^2$, {\em which play the role of the observation scale} \cite{Altarelli2013,DEUR2016}.

In theoretical models, the straightforward introduction of a cutoff momentum $\Lambda$ as the scale of observation 
is not always successful. A physical theory should be Lorentz invariant, should provide energy and momentum conservation, and may have gauge and other symmetries. The use of the truncated fields 
$$
\phi^{<}(x) := \int_{|k|<\Lambda} e^{-\imath k x} \tilde{\phi}(k) \dk{k}{d}
$$
may destroy the symmetries. In the limiting case of $\Lambda\to\infty$, this returns to the standard Fourier 
transform, making some of the Green functions $\bra \phi(x_1)\ldots \phi(x_n)\ket$ infinite and the theory meaningless. A practical solution of this problem was found in the renormalization group (RG) method \cite{Collins1984},
first discovered  in quantum electrodynamics \cite{SP1953}. The bare charges and the bare fields of the theory are then 
renormalized to some ''physical''   charges and fields, the Green functions for which become finite. The price to be 
paid for it is the appearance in the theory of some new normalization scale $\mu^2$. The comparison of the model 
prediction to the experimental observations now requires the use of two scale parameters ($Q^2,\mu^2$) \cite{Collins1984}.

A significant disadvantage of the RG method is that in renormalized theories, we are often doomed to ignore the 
finite parts of the Feynman graphs.  
The solution of the divergences problem may be the change of the functional space to the space of functions that 
explicitly depend on both the position  and the resolution  -- the scale of observation. The Green functions for such fields $\bra \phi_{a_1}(x_1)\ldots\phi_{a_n}(x_n)\ket$ can be made finite by construction 
under certain causality conditions \cite{CC2005,AK2013}. 

The introduction of resolution into the definition of 
the field function has a clear physical interpretation. 
If the particle, described by the field $\phi$, has been 
initially prepared in the interval 
$(x-\frac{\Delta x}{2},x+\frac{\Delta x}{2})$, the probability of 
registering this particle in this interval is generally less than unity, because 
the probability of registration depends on the strength of interaction 
and on the ratio of typical scales of the measured particle and the measuring 
equipment. The maximum probability of registering an object of 
typical scale $\Delta x$ by equipment with typical resolution $a$
 is achieved when these two parameters are comparable. For this reason, 
the probability of registering an electron by visual-range photon scattering 
is much higher than  that by long radio-frequency waves. As a  
mathematical generalization, we should say that if a set of measuring equipment  
with a given spatial resolution $a$ fails to register an object, prepared 
on a spatial interval of width $\Delta x$ with certainty, 
then tuning the equipment to {\em all} possible resolutions $a'$ would 
lead to the registration -- 
$
\int |\phi_a(x)|^2 d\mu(a,x) = 1,
$
where $d\mu(a,x)$ is some measure, that depends on resolution $a$.
 This certifies the fact of the existence 
of the measured object.

A straightforward way to construct a space of scale-dependent functions is to use a projection of 
local fields $\phi(x) \in \mathrm{L^2}(\R^d)$ onto some basic function $\chi(x)$ with good localization properties, in 
both the position and momentum spaces, and scaled to a typical window width of size $a$. This can be achieved by a {\em continuous wavelet transform} \cite{Daub10}.

\subsection{Continuous wavelet transform}
Let $\cH$ be a Hilbert space of states for a quantum field $|\phi\ket$. 
Let $G$ be a locally compact Lie group acting transitively on $\cH$, 
with $d\mu(\nu),\nu\in G$ being a left-invariant measure on $G$. Then, 
any $|\phi\ket \in \cH$ can be decomposed with respect to 
a representation $\Omega(\nu)$ of $G$ in $\cH$ \cite{Carey1976,DM1976}:
\begin{equation}
|\phi\ket= \frac{1}{C_\chi}\int_G \Omega(\nu)|\chi\ket d\mu(\nu)\bra \chi|\Omega^\dagger(\nu)|\phi\ket, \label{gwl} 
\end{equation} 
where $|\chi\ket \in \cH$ is referred to as a {\em mother wavelet}, satisfying the admissibility condition 
$$
C_\chi = \frac{1}{\| \chi \|^2} \int_G |\bra \chi| \Omega(\nu)|\chi \ket |^2 
d\mu(\nu)
<\infty. 
$$
The coefficients $\bra \chi|\Omega^\dagger(\nu)|\phi\ket$ are referred to as 
wavelet coefficients. 
Wavelet coefficients can be used in quantum mechanics in the same spirit 
as the coherent states are used \cite{DGM1986,KlaStre91}.

If the group $G$ is Abelian, the wavelet transform [Eq.\eqref{gwl}] with 
$G:x'=x+b'$ is the  Fourier transform. 
Next to the Abelian group is the group of the affine transformations 
of the Euclidean space $\R^d$:
\begin{equation}
G: x' = a R(\theta)x + b, x,b \in \R^d, a \in \R_+, \theta \in SO(d), \label{ag1}
\end{equation} 
where $R(\theta)$ is the $SO(d)$ rotation matrix.
Here we define the representation of the affine transform [Eq.\eqref{ag1}] with 
respect to the mother wavelet $\chi(x)$ as follows:
\begin{equation}
U(a,b,\theta) \chi(x) = \frac{1}{a^d} \chi \left(R^{-1}(\theta)\frac{x-b}{a} \right).
\end{equation}

Thus the wavelet coefficients of the function $\phi(x) \in L^2(\R^d)$ with 
respect to the mother wavelet $\chi(x)$ in Euclidean space $\R^d$ can be written 
as 
\begin{equation}
\phi_{a,\theta}(b) = \int_{\R^d} \frac{1}{a^d} \overline{\chi \left(R^{-1}(\theta)\frac{x-b}{a} \right) }\phi(x) d^dx. \label{dwtrd}
\end{equation} 
The wavelet coefficients \eqref{dwtrd} represent the result of the measurement 
of function $\phi(x)$ at the point $b$ at the scale $a$ with an aperture 
function $\chi$ rotated by the angle(s) $\theta$ \cite{PhysRevLett.64.745}.
The function $\phi(x)$ can be reconstructed from its wavelet coefficients 
[Eq.\eqref{dwtrd}] using the formula \eqref{gwl}:
\begin{equation}
\phi(x) = \frac{1}{C_\chi} \int \frac{1}{a^d} \chi\left(R^{-1}(\theta)\frac{x-b}{a}\right) \phi_{a\theta}(b) \frac{dad^db}{a} d\mu(\theta), \label{iwt}
\end{equation}
where $d\mu(\theta)$ is the left-invariant measure on the $SO(d)$ rotation group, usually written in terms of the 
Euler angles: $$d\mu(\theta) = 2\pi \prod_{k=1}^{d-2} \int_0^\pi \sin^k \theta_k d\theta_k.$$
The normalization 
constant
$C_\chi$ is readily evaluated using the Fourier transform.
In what follows, I assume  isotropic wavelets and omit the angle variable $\theta$. This means that the mother wavelet $\chi$ is assumed to be invariant under $SO(d)$ rotations. This is quite common for the problems with no preferable directions. For isotropic 
wavelets, 
\begin{equation}
C_\chi = \int_0^\infty |\tilde \chi(ak)|^2\frac{da}{a}
= \int |\tilde \chi(k)|^2 \frac{d^dk}{S_{d}|k|^d} < \infty,
\label{adcfi}
\end{equation}
where $S_d = \frac{2 \pi^{d/2}}{\Gamma(d/2)}$ is the area of the unit sphere 
in $\R^d$, with $\Gamma(x)$ being  Euler's gamma function. A tilde denotes the Fourier transform: $\tilde{\chi}(k) = \int_{\R^d} e^{\imath k x} \chi(x) d^d x$.

If the standard quantum field theory defines the field 
function $\phi(x)$ as a scalar product of the state vector of the system 
 and the state vector corresponding to the localization at the point $x$:
$
\phi(x) \equiv \bra x | \phi \ket,
$
the modified theory \cite{AltSIGMA07,Altaisky2010PRD} should respect the resolution of the measuring equipment. Namely, we define the 
{\em resolution-dependent fields} 
\begin{equation}
\phi_{a}(x) \equiv \bra x,a; \chi|\phi\ket,\label{sdf}
\end{equation}
also referred to as the scale components of $\phi$,
where $\bra x,  a; \chi|$ is the bra-vector corresponding to localization 
of the measuring device around the point $x$ with the spatial resolution $a$;
in optics $\chi$ labels the apparatus function of the equipment, an {\em aperture} function
\cite{PhysRevLett.64.745}. 

In QED, the common calculation techniques are based on the basis of plane waves. However, the basis of plane waves is not obligatory. For instance, if the inverse size of a QED microcavity is compared to the energy of an interlevel transition of an atom, or a quantum dot inside the cavity, we can (at least in principle) avoid the use of plane waves and use some other functions to estimate the dependence of vacuum energy effects on the size and shape of the cavity. In this sense, the mother wavelet can be referred to as an aperture function.
In QCD, all measuring equipment is far removed from the collision domain, and the 
approximation of plane waves may be most simple technically, but it is not justified unambiguously: some other sets of functions may be used as well. Discrete wavelet basis, for instance, has been already used for common QCD models in Ref. \cite{Federbush1995}.
The field theory of extended objects with the basis $\chi$  defined on the spin variables was considered in Refs. \cite{GS2009,Varlamov:2012}.

The goal of the present paper is to study the scale dependence of the running coupling constant in non-Abelian gauge theory constructed directly on scale-dependent fields. 
Assuming the mother  
wavelet $\chi$ to be isotropic, we drop the angle argument $\theta$ in Eq. \eqref{iwt} and perform all calculations in Euclidean 
space. 

The interpretation of the real experimental results in terms of the wave packet $\chi$ is a nontrivial problem to be 
of special concern in future. It can be addressed by constructing wavelets in the Minkowski space and by analytic continuation from the Euclidean space to the Minkowski space \cite{PG2012,AK2013}.

 For the same reason, I also do not consider 
here the quantization of scale-dependent fields, which was addressed elsewhere \cite{BP2013,AK2013,AK2016IJTP}.
A prospective way to do this, as suggested in Refs.\cite{AK2013,AK2013iv}, is the use of light-cone coordinates 
\cite{BT2008,Polyzou2020}.
With these remarks we can understand the physically measured fields, at least in local theories like QED and the $\varphi^4$ model, as the integrals over all scale components from the measurement scale ($A$) to infinity: 
$$
\phi^{(A)}(x) =\frac{1}{C_\chi}\int_{a\ge A} 
\bra x|\chi;a,b\ket d\mu(a,b)\bra \chi;a,b|\phi\ket.
$$
The limit of an infinite resolution ($A\to0$) certainly drives us back to the known 
divergent theories. 

\subsection{An example of scalar field theory}
To illustrate the wavelet method, following the previous papers \cite{Altaisky2010PRD,Altaisky2016PRD}, I   
start with the phenomenological model of a scalar field with nonlinear self-interaction $\phi^4(x)$, described by 
the Euclidean action functional 
\begin{equation}
S_E[\phi] = \int d^d x \bigl[ \frac{1}{2}(\partial\phi)^2 + \frac{m^2}{2}\phi^2 + \frac{\lambda}{4!}\phi^4
\bigr].
\label{se4:eq}
\end{equation}
This model is an extrapolation of a classical interacting spin model to the continual limit \cite{GJ1981}. Known as the Ginzburg-Landau 
model \cite{GL1950}, it describes phase transitions in superconductors and other  magnetic systems fairly well, but it produces divergences when the 
correlation functions 
\begin{equation}
G^{(n)}(x_1,\ldots,x_n) = 
\left. { \frac{\delta^n\ln Z[J]}{\delta J(x_1) \ldots \delta J(x_n)}
}\right|_{J=0}
\label{cgf}
\end{equation} 
are evaluated from the generating functional 
\begin{equation}
Z[J] = \cN \int e^{-S_E[\phi]+ \int J(x) \phi(x) d^dx} \cD \phi
\end{equation}
[where $J(x)$ is a formal source, used to calculate the Green functions, and $\cN$ is a formal normalization 
constant of the Feynman integral]  
by perturbation expansion; see, e.g., Ref.\cite{ZJ1999}.

The parameter $\lambda$ in the action functional [Eq.\eqref{se4:eq}] is a phenomenological coupling constant, which knows nothing about 
the scale of observation, and becomes the running coupling constant only because of renormalization or cutoff 
introduction. The straightforward way to introduce the scale dependence into the model [Eq.\eqref{se4:eq}] is to express 
the local field $\phi(x)$ in terms of its scale components $\phi_a(b)$ using the inverse wavelet transform [Eq.\eqref{iwt}]:
\begin{equation}
\phi(x) = \frac{1}{C_\chi} \int \frac{1}{a^d} \chi\left(\frac{x-b}{a}\right) \phi_{a}(b) \frac{dad^db}{a}. \label{iiwt}
\end{equation}
This leads to the generating functional for scale-dependent fields: 
\begin{widetext}
\begin{align} \nonumber 
Z_W[J_a] &=&\cN \int \cD\phi_a(x) \exp \Bigl[ -\frac{1}{2}\int \phi_{a_1}(x_1) D(a_1,a_2,x_1-x_2) \phi_{a_2}(x_2)
\frac{da_1d^dx_1}{C_\chi a_1}\frac{da_2d^dx_2}{C_\chi a_2}  \\
&-&\frac{\lambda}{4!}
\int V_{x_1,\ldots,x_4}^{a_1,\ldots,a_4} \phi_{a_1}(x_1)\cdots\phi_{a_4}(x_4)
\frac{da_1 d^dx_1}{C_\chi a_1} \frac{da_2 d^dx_2}{C_\chi a_2} \frac{da_3 d^dx_3}{C_\chi a_3} \frac{da_4 d^dx_4}{C_\chi a_4} 
+ \int J_a(x)\phi_a(x)\frac{dad^dx}{C_\chi a}\Bigr], \label{gfw}
\end{align}
\end{widetext}
where $D(a_1,a_2,x_1-x_2)$ is the wavelet transform of the ordinary propagator, and $\cN$ is a formal normalization constant.

The functional \eqref{gfw} -- if integrated over all scale arguments in infinite limits $\int_0^\infty \frac{da_i}{a_i}$ -- will certainly drive us back to the known divergent theory. All scale-dependent fields [$\phi_a(x)$] in 
Eq.\eqref{gfw} still interact with each other with the same coupling constant $\lambda$, but their interaction is now 
modulated  by the wavelet factor $V_{x_1x_2x_3x_4}^{a_1a_2a_3a_4}$, which is the Fourier transform of 
$\prod_{i=1}^4 \tilde{\chi}(a_ik_i)$.
 In coordinate form, for the $\frac{\lambda}{N!}\phi^N$ interaction, these coefficients, calculated with the above mentioned first derivative of the Gaussian taken as the mother wavelet, have the 
form 
\begin{align*}
V_{b_1,\ldots, b_N}^{a_1,\ldots,a_N} = (-1)^N\left(\frac{2\pi}{\zeta}\right)^\frac{d}{2} 
\exp\left( 
-\frac{1}{2} \sum_{k=1}^N \left(\frac{b_k}{a_k} \right)^2 
\right) \times \\ 
\times \exp\left( \frac{\xi^2}{2\zeta}\right) 
\prod_{i=1}^N \frac{1}{a_i^{d+1}} \left(\frac{\xi}{\zeta}-b_i \right), \\
\zeta \equiv \sum_{k=1}^N \frac{1}{a_k^2}, 
\xi \equiv \sum_{k=1}^N \frac{b_k}{a_k^2},
\end{align*}
where $d$ is the space dimension, and $1/\sqrt{\zeta}$ is a kind of weighted scale.

For Feynman diagram expansion, the substitution of the fields by Eq.\eqref{iiwt} is 
naturally performed in the Fourier representation 
\begin{align*}
\phi(x) &=& \frac{1}{C_\chi} \int_0^\infty \frac{da}{a} \int \dk{k}{d} e^{-\imath k x}
\tilde \chi(ak) \tilde \phi_a(k), 
\\  
\tilde\phi_a(k) &=& \overline{\tilde \chi(ak)}\tilde\phi(k) .
\end{align*}
Doing so, we have the following modification of the Feynman diagram technique
\cite{Alt2002G24J}:
\begin{itemize}\itemsep=0pt
\item Each field $\tilde\phi(k)$ is substituted by the scale component
$\tilde\phi(k)\to\tilde\phi_a(k) = \overline{\tilde \chi(ak)}\tilde\phi(k)$.
\item Each integration in the momentum variable is accompanied by the corresponding 
scale integration
\[
 \dk{k}{d} \to  \dk{k}{d} \frac{da}{a} \frac{1}{C_\chi}.
 \]
\item Each interaction vertex is substituted by its wavelet transform; 
for the $N$th power interaction vertex, this gives multiplication 
by the factor 
$\displaystyle{\prod_{i=1}^N \tilde \chi(a_ik_i)}$.
\end{itemize}
According to these rules, the bare Green function in wavelet representation 
takes the form
$$
G^{(2)}_0(a_1,a_2,p) = \frac{\tilde \chi(a_1p)\tilde \chi(-a_2p)}{p^2+m^2}.
$$ 
The finiteness of the loop integrals is provided by the following rule:
{\em There should be no scales $a_i$ in internal lines smaller than the minimal scale 
of all external lines} \cite{Alt2002G24J,Altaisky2010PRD}. Therefore, the integration in $a_i$ variables is performed from 
the minimal scale of all external lines up to infinity.

For a theory with local $\phi^N(x)$ interaction the presence of two conjugated factors $\tilde{\chi}(ak)$ and 
$\overline{\tilde{\chi}(ak)}$ on each diagram line, connected to the interaction vertex, simply means that each internal 
line of the Feynman diagram carrying momentum $k$ is supplied by the cutoff factor $f^2(Ak)$, where 
\begin{equation}
f(x) := \frac{1}{C_\chi}\int_x^\infty |\tilde \chi(a)|^2\frac{da}{a},
\quad f(0)=1, \label{cutf1}
\end{equation} 
where $A$ is the minimal scale of all external lines of this diagram. This factor automatically suppresses all 
UV divergences.

The admissibility condition \eqref{adcfi} for the mother wavelet $\chi$ is rather loose. At  best, $\chi(x)$ would 
be the aperture function of the measuring device \cite{PhysRevLett.64.745}. In practice, 
any well-localized function with $\tilde{\chi}(0)=0$ will suit. For analytical calculations, the mother 
wavelet should be easy to integrate, and for this reason, as in previous papers \cite{Altaisky2010PRD,AK2013,Altaisky2016PRD}, we choose the derivative of the 
Gaussian function as a mother wavelet:
\begin{equation}
\tilde{\chi}(k) = -\imath k e^{-\frac{k^2}{2}}. \label{g1f:eq}
\end{equation}
This gives $C_\chi=\frac{1}{2}$ and provides the exponential cutoff factor [Eq.\eqref{cutf1}]: $f(x)=e^{-x^2}$.

As usual in functional renormalization group technique \cite{Wetterich1993}, we can introduce the effective action 
functional: 
$$
\Gamma[\phi]=-W[J]+ \bra J \phi \ket,
$$
the functional derivatives of which are the vertex functions:
\begin{widetext}
$$
\Gamma_{(A)}[\phi_a] = \Gamma_{(A)}^{(0)} + \sum_{n=1}^\infty \int  \frac{1}{C_\chi^n}
\Gamma_{(A)}^{(n)}(a_1,b_1,\ldots,a_n,b_n) \phi_{a_1}(b_1)
\ldots \phi_{a_n}(b_n) \frac{da_1d^db_1}{a_1}
\ldots \frac{da_nd^db_n}{a_n}.
$$
\end{widetext}
The subscript $(A)$ indicates the presence in the theory of 
minimal scale -- the observation scale. 

Let us consider the one-loop vertex function $\Gamma^{(4)}_{(A)}$ in the scale-dependent $\phi^4$ model with 
the mother  wavelet \eqref{g1f:eq} \cite{Altaisky2016PRD}. The  $\Gamma^{(4)}_{(A)}$ contribution to the effective action is shown in diagram \eqref{G4:fde}:
\begin{equation}
\Gamma^{(4)} = -
\begin{tikzpicture}[baseline=(a)]
\begin{feynman}[inline=(a)]
    \vertex [dot] (a);
    \vertex [above=1cm of a] (i1){\(1\)};
    \vertex [left =1cm of a] (i2){\(2\)};
    \vertex [right=1cm of a] (i3){\(3\)};
    \vertex [below=1cm of a] (i4){\(4\)};
    \diagram*{  
    {(i1),(i2),(i3),(i4)} -- [gluon] (a),
    };
\end{feynman}      
\end{tikzpicture}
  -\frac{3}{2}
  \begin{tikzpicture}[baseline=(e)]
  \begin{feynman}[horizontal = (e) to (f)]
  \vertex [dot] (e);
  \vertex [dot,right=1.5cm of e] (f);
  \vertex [above left = 1cm of e] (i1) {\(1\)};
  \vertex [below left = 1cm of e] (i2) {\(2\)};
  \vertex [above right = 1cm of f] (i3) {\(3\)};
  \vertex [below right = 1cm of f] (i4) {\(4\)};
  \diagram*{
  {(i1),(i2)} -- [scalar] (e),
  {(i3),(i4)} -- [scalar] (f),
  (e) -- [scalar,half right, momentum=\(q\)] (f),
  (e) -- [scalar,half left] (f),
    };
  \end{feynman} 
  \end{tikzpicture}  
  \label{G4:fde}
\end{equation}
Each vertex of the Feynman diagram corresponds to $-\lambda$, and each external line of the 1PI diagram contains the wavelet factor $\tilde\chi(a_ik_i)$, hence 
\begin{equation}
\frac{\Gamma^{(4)}_{(A)}}{\tilde{\chi}(a_1p_1)\tilde{\chi}(a_2p_2)\tilde{\chi}(a_3p_3)\tilde{\chi}(a_4p_4) } = \lambda -\frac{3}{2}\lambda^2 X^d(A). \label{g4l1}
\end{equation}
The value of the one-loop integral
\begin{equation}
X^d(A) = \int \frac{d^dq}{(2\pi)^d}
\frac{f^2(qA)f^2((q-s)A)}{\left[ q^2+m^2\right]\left[ (q-s)^2+m^2\right] },
\label{li1}
\end{equation}
where $s\!=\!p_1\!+\!p_2$ and $A=\min(a_1,a_2,a_3,a_4)$, 
depends on the mother  wavelet $\chi$ by means of the cutoff function $f(x)$. 
The integral in Eq.\eqref{li1} with the Gaussian cutoff function [Eq.\eqref{cutf1}] can be easily evaluated.
In physical dimension $d=4$ in the limit $s^2 \gg 4m^2$, this gives \cite{Altaisky2010PRD}
\begin{align}\nonumber 
\lim_{s^2\gg 4m^2} X^4(\alpha^2) &=& \frac{e^{-2\alpha^2}}{16\pi^2\alpha^2} 
\bigl[e^{\alpha^2}-1 
- \alpha^2e^{2\alpha^2}\Ei_1(\alpha^2) \\
&+& 2\alpha^2e^{2\alpha^2}\Ei_1(2\alpha^2)
  \bigr], \label{x4l1}
\end{align}
where $\alpha = As$ is a dimensionless scale, and $$\Ei_1(x)\equiv \int_1^\infty \frac{e^{-xt}}{t}dt$$ is the exponential integral of th first type. All integrals are finite now, and the coupling constant becomes {\em running}, 
$\lambda = \lambda(\alpha^2)$, only because of its dependence on the dimensionless observation scale $\alpha$:
\begin{equation}
\frac{\partial \lambda}{\partial\mu} = 3\lambda^2\alpha^2 \frac{\partial X^4}{\partial\alpha^2} = \frac{3\lambda^2}{16\pi^2} \frac{2\alpha^2+1-e^{\alpha^2}}{\alpha^2}
e^{-2\alpha^2}, \label{b1}
\end{equation}
where $\mu = -\ln A + const$. The dimensionless scale variable $\alpha$ is the product of the observation scale $A$ and the total momentum $s$. 
The analogue of Eq. \eqref{x4l1} in standard field theory subjected to cutoff at momentum $\Lambda$ is
$$
X_\Lambda^d = \int_{|q|\le \Lambda} \frac{d^dq}{(2\pi)^d} \frac{1}{(q^2+m^2)((q-s)^2+m^2)}.
$$ 
Symmetrizing the latter equation in the loop momenta $q \to q + s/2$, we get, 
in the same limit of $s^2 \gg 4 m^2$ and the dimension $d=4$: 
\begin{equation}
X_\Lambda^4 = \frac{1}{16\pi^2} \ln\left(4 \left(\frac{\Lambda}{s} \right)^2 + 1 \right) \label{x4lambda}
\end{equation}
We can compare Eq.\eqref{x4lambda} to Eq. \eqref{x4l1} by setting $\frac{\Lambda^2}{s^2}=1/r$ in momentum space, and $\alpha^2 = r$ in wavelet space. Graphs showing the dependence of the one-loop contribution to the $\phi^4$ vertex as a function of scale for both the standard [Eq.\eqref{x4lambda}] and the wavelet-based  [Eq.\eqref{x4l1}] formalisms are presented in Fig.~\ref{x4r:pic} below.
\begin{figure}[h]
\centering \includegraphics[width=7cm]{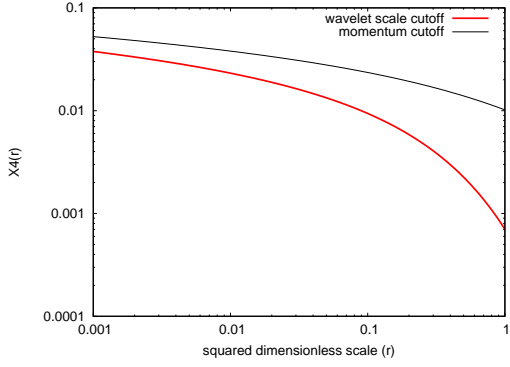}
\caption{Plot of the one-loop contribution  to the $\phi^4$ vertex, calculated  with the first derivative of the Gaussian as the mother  wavelet as a function of the squared observation scale $r=A^2 s^2$, compared to that calculated with the standard cutoff at the cutoff momentum $\Lambda = A^{-1}$ in Euclidean $d=4$ dimensions.}
\label{x4r:pic}
\end{figure}
The slopes of both curves in UV limit ($A\to0$) are the same: $\frac{\partial \lambda}{\partial\mu} = \frac{3\lambda^2}{16\pi^2}$.

The running coupling constant $\lambda(\alpha^2)$ can be understood as the coupling that folds into its running all quantum effects characterized by a scale larger than the observation scale $A$. 

For small $\alpha$, Eq.\eqref{b1} tends to the known result. 
This is because we have started with the local Ginzburg-Landau theory, where the fluctuations of all scales 
interact with each other, with the interaction of neighbouring scales being most important; see, e.g.\cite{WK1974}
for an excellent discussion of the underlying physics. 

\subsection{QED: wavelet regularization of a local gauge theory}
Quantum electrodynamics is the simplest gauge theory of the type given in Eq. \eqref{de1}, with the gauge 
group being the Abelian group $U(1)$:
\begin{equation}
\psi(x) \to e^{-\imath e \Lambda(x)} \psi(x). \label{B:eq}
\end{equation}
The transformation of the gauge field -- the electromagnetic field -- is the gradient 
transformation:  
\begin{equation}
A_\mu(x) \to A_\mu(x) + \partial_\mu \Lambda(x). 
\label{gtu1:eq}
\end{equation}
In view of the linearity of the wavelet transform, Eq. \eqref{gtu1:eq} keeps the same form for all 
scale components of the gauge field $A_{\mu, a}(x)$ -- in contrast to the matter field transformation [Eq.\eqref{B:eq}], 
which is nonlinear -- and thus, the gauge transform of the matter fields in a local gauge theory is not the change of 
all scale components $\psi_a(x)$ by the {\em same} phase.

The Euclidean QED Langangian is:
\begin{align}
L &=& \bar\psi(x)(\slashed{D}+ \imath m)\psi(x) 
+ \frac{1}{4} F_{\mu\nu}F_{\mu\nu}
+ \underbrace{
\frac{1}{2\alpha} (\partial_\mu A_\mu)^2
}_{\hbox{gauge fixing}}, \label{gau1l}\\ \nonumber
&& D_\mu \equiv \partial_\mu +\imath e A_\mu(x), \hbox{with\ }  F_{\mu\nu}=\partial_\mu A_\nu - \partial_\nu A_\mu
\end{align}
which is the field strength tensor of the electromagnetic field $A_\mu(x)$. (The slashed vectors denote the convolution  
with the Dirac $\gamma$ matrices: $\slashed{D}\equiv \gamma_\mu D_\mu$.)

The wavelet regularization technique works for QED in the same way as it does for the above considered scalar field 
theory. This means that each line of the Feynman diagram carrying momentum $p$ acquires a cutoff factor $f^2(Ap)$. 

In this way, in one-loop approximation, we get the electron self-energy, shown in  Fig.~\ref{sed:pic}:
\begin{figure}
\feynmandiagram[small,horizontal= a to b] {
i1[particle=a] -- [fermion,momentum=\(p\)] a -- [photon, half left,momentum=\(q+p/2\)]  b --[fermion,momentum=\(p\)] o1[particle=a'],
a -- [fermion] b
};
\caption{Electron self-energy diagram in scale-dependent QED \label{sed:pic}.}
\end{figure}
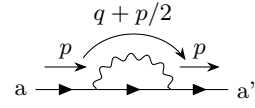
\begin{equation}
\frac{\Sigma^{(A)}(p)}{\tilde \chi(a p) \tilde \chi(-a' p)} = -\imath e^2 \int \dk{q}{4}  
\frac{F_A(p,q)
\gamma_\mu 
 \left[\frac{\slashed{p}}{2}-\slashed{q}-m \right] \gamma_\mu
 }
 {\left[
\left(\frac{p}{2}-q \right)^2+m^2\right]
 \left[\frac{p}{2}+q \right]^2
 },
\end{equation}
where 
$$
F_A(p,q) \equiv f^2\left(A(\frac{p}{2}+q) \right)  f^2\left(A(\frac{p}{2}-q) \right)=e^{-A^2p^2-4A^2q^2}
$$
is the product of the wavelet cutoff factors, and 
$A=\min(a,a')$ is the minimal scale of two external lines of 
the diagram Fig.~\ref{sed:pic}.

\begin{figure}
\feynmandiagram[small,horizontal= a to b] {
i1[particle=a] -- [photon,momentum=\(p\)] a -- [fermion, half left,momentum=\(q+p/2\)]  b --[photon,momentum=\(p\)] o1[particle=a'],
b -- [fermion,half left] a
};
\caption{Vacuum polarization diagram in scale-dependent QED \label{vpd:pic}.}
\end{figure}
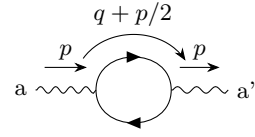

\begin{figure}
\begin{tikzpicture}
\begin{feynman}[vertical= e to f] 
   \vertex (e);
   \vertex [below = 1cm of e] (f) {\(r\)};
   \vertex [above right = 1.5cm of e] (b);
   \vertex [above left  = 1.5cm of e] (c);
   \vertex [above right = 1cm of b] (a) {\(a\)};
   \vertex [above left  = 1cm of c] (c1) {\(a'\)};
   \diagram*{ 
    (a)  -- [fermion,momentum=\(k\)] (b) -- [fermion,momentum=\(k-f\)] (e) -- [fermion] (c) 
    -- [fermion,momentum=\(p\)] (c1),
    (b) -- [photon,momentum=\(f\)] (c),
    (e) -- [photon,momentum=\(q\)] (f),
    }; 
\end{feynman}    
\end{tikzpicture}
\caption{One-loop vertex function in scale-dependent QED \label{ep1:pic}}
\end{figure}
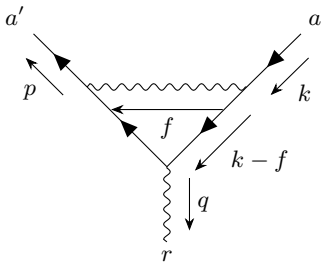

Similarly, for the vacuum polarization diagram of QED, shown in Fig.~\ref{vpd:pic}, we get \cite{AltSIGMA07}
\begin{align}\nonumber 
\frac{\Pi_{\mu\nu}^{(A)}(p)}{\tilde \chi (ap) \tilde \chi(-a'p)}&=& -e^2 \int \dk{q}{4} F_A(p,q) \times \\
 &\times& \frac{\mathrm{Tr} (\gamma_\mu (\slashed{q}+ \slashed{p}/2  -m)\gamma_\nu (\slashed{q} -\slashed{p}/2 - m))}{\left[(q+p/2)^2+m^2\right]\left[(q-p/2)^2+m^2\right]} 
.
\label{padef}
\end{align}
The electron-photon interaction vertex, in one-loop approximation, with the photon propagator taken in the Feynman gauge,
gives the equation 
\begin{align}\nonumber 
\frac{\Gamma_{\mu,r}^{(A)}}{\tilde{\chi}(-pa') \tilde{\chi}(-qr) \tilde{\chi}(ka)}  = e^2 
\int  \dk{f}{4} \gamma_\alpha 
\frac{\slashed{p}-\slashed{f}-m}{(p-f)^2+m^2}  \\
\times \gamma_\mu   \frac{\slashed{k}-\slashed{f}-m}{(k-f)^2+m^2}    \gamma_\alpha \frac{1}{f^2} f^2(A(p-f)) f^2(A(k-f)) f^2(Af). \label{l1v}
\end{align} 
The vertex function [Eq.\eqref{l1v}] and the inverse propagator 
are related by the Ward-Takahashi identities, which are wavelet 
transforms of corresponding identities of the ordinary local gauge 
theory \citep{AA2009,AK2013}. 
The detailed one-loop calculations, except for the contribution to the vertex, can be found in Ref.\cite{AK2013}. As for 
the vertex contribution [Eq.\eqref{l1v}], shown in Fig.~\ref{ep1:pic}, the calculation is rather cumbersome, but  it can be done numerically.   

\section{Gauge invariance for scale-dependent fields \label{gis:sec}}
For a non-Abelian gauge theory both terms in the gauge field transformation [Eq.\eqref{at}]
are nonlinear. The wavelet transform [Eq.\eqref{iiwt}] can  hardly be applied to the theory without violation of local gauge invariance.
An attempt  to use wavelets for  gauge theories, QED and QCD, was undertaken for the first time by P.Federbush \cite{Federbush1995} 
in a form of discrete wavelet transform. 
Later, it was extended by using the wavelet transform in lattice simulations and theoretical studies of related 
problems \cite{HS1995,Battle1999,Polyzou2017}.
The consideration was restricted to axial gauge and a special type of divergency-free wavelets in four dimensions. The context of that application was the localization of the wavelet basis, which may be beneficial 
for numerical simulation, but is not tailored for analytical studies, and does not link the gauge invariance to the dependence on scale. 
The discrete wavelet transform approach to different quantum field theory problems has been further developed in Hamiltonian formalism, but for scalar theories with local interaction \cite{Brennen2015,Polyzou2017}.

Now it is a point to think of how we can build a gauge-invariant theory of fields that depend on both the position ($x$) and the resolution 
($a$). To do this, we recall that the free fermion action [Eq.\eqref{de}] can be considered as a matrix element of the 
Dirac operator:
\begin{equation}
S_E = \bra \psi |\gamma_\mu\partial_\mu + \imath m|\psi\ket.
\label{sme}
\end{equation} 
Assuming a scalar product $\bra\cdot|\cdot\ket$ in a general Hilbert space $\cH$, in accordance with the original Dirac's formulation of quantum field theory \cite{DiracPQM4}, we can insert arbitrary partitions of unity 
$
\hat{1}=\sum_c |c\ket\bra c|
$
into Eq. \eqref{sme}, so that 
$$
S_E = \sum_{c,c'} \bra\psi |c\ket 
\bra c| 
\gamma_\mu\partial_\mu + \imath m |c'\ket
\bra c'|\psi\ket.
$$
An important type of the unity partition in Hilbert space $\cH$ is a 
unity partition related to the generalized wavelet transform [Eq.\eqref{gwl}]:
\begin{equation}
\hat{1} = \int_{G} \Omega(\nu) |\chi\ket \frac{d\mu_L(\nu)}{C_\chi}\bra \chi|\Omega^\dagger(\nu).
\end{equation}

Our main criterion for this choice is to find a group $G$ which pertains to the physics of quantum measurement and provides the fields defined 
on finite domains rather than points. The group that can leverage this task is a group of affine transformations: 
\begin{equation}
G:x' = a x + b, a \in \R_+, x',x,b \in \R^d.
\label{Ag}
\end{equation} 
Following Refs.\cite{Altaisky2010PRD,AK2013}, we consider an isotropic theory.
The representation of the affine group [Eq.\eqref{Ag}] in  $L^2(\R^d)$ is chosen as 
\begin{equation}
[\Omega(a,b)\chi](x) := \frac{1}{a^d}\chi \left(
\frac{x-b}{a}
\right),
\end{equation}
and the left-invariant Haar measure is  
\begin{equation}
d\mu_L(a,b)=\frac{da d^db}{a}. \label{hm}
\end{equation}
In view of the linearity of the wavelet transform 
\begin{equation}
\psi(x) \to \psi_a(b) = \int_{\R^d} \frac{1}{a^d}\bar{\chi}
\left(\frac{x-b}{a}
\right) \psi(x) d^dx,
\label{cwt}
\end{equation}
the action on the affine group [Eq.\eqref{Ag}] keeps the same form 
as the action of the genuine theory [Eq.\eqref{sme}]. Thus, we get the action functional for the fields $\psi_a(b)$ defined on the affine group:
\begin{align}\nonumber 
S_E &=& \frac{1}{C_\chi} \int_{\R_+\otimes\R^d} \Bigl[
\bar{\psi}_a(b)\gamma_\mu \partial_\mu \psi_a(b) \\
&+& \imath m \bar{\psi}_a(b) \psi_a(b)
\Bigr] \frac{da d^db}{a}, \label{sag}
\end{align}  
where the derivatives $\partial_\mu$ are now taken with respect to spatial variables $b_\mu$. 
The meaning of the 
representation Eq.\eqref{sag} is that the action functional  
 is now a sum of {\em independent} scale components $S_E = \int S(a)\frac{da}{a}$, with no interaction between the scales. 

Starting from the locally gauge invariant action $S_E=\int d^4x \bar\psi (\slashed{D} + \imath m)\psi$ we  destroy such independence by 
the cubic term $\bar{\psi}\slashed{A} \psi$, which yields 
cross-scale terms. However, knowing nothing about the {\em point-dependent} gauge fields $A_\mu(x)$ at this stage, we should  certainly ask the question of how one can make the theory of Eq.\eqref{sag} invariant with respect to a phase transformation defined locally on the affine group: 
\begin{equation}
U_a(b) = \exp\left( \imath \sum_A w^A_a(b)T^A\right)? \label{grot}
\end{equation}

Since the action [Eq.\eqref{sag}], for each fixed value of the scale $a$, has exactly the same form as the standard action [Eq.\eqref{de}], we can introduce 
the invariance with respect to local phase transformation separately at 
each scale by changing the derivative $\partial_\mu\equiv\frac{\partial}{\partial b_\mu}$ into the covariant derivative
\begin{equation}
D_{\mu,a}= \partial_\mu + \imath A_{\mu,a}(b), \label{cds} 
\end{equation}
with the gauge transformation law for the scale-dependent gauge field 
$A_{\mu,a}(b)=\sum_A A_{\mu,a}^A(b)T^A$ identical to Eq.\eqref{at}:
$$
A_{\mu,a}'(b) = U_a(b) A_{\mu,a}(b) U^\dagger_a(b) + \imath \left(\partial_\mu U_a(b) \right) U^\dagger_a (b). $$
Similarly, for the field strength tensor  and for the Yang-Mills Lagrangian:
\begin{equation}
F_{\mu\nu,a}=-\imath [D_{\mu,a},D_{\nu,a}],\ L_a^{YM} = \frac{1}{2g^2}\Tr ( F_{\mu\nu,a}F_{\mu\nu,a}).
\end{equation}
Assuming the formal coupling constant of the gauge field $A_{\mu,a}(b)$ to be dependent on scale only, we can rewrite the covariant derivative by changing $A_{\mu,a}(b)$ to $g(a)A_{\mu,a}(b)$:  
\begin{equation}
D_{\mu,a}= \partial_\mu + \imath g A_{\mu,a}(b). \label{cdsa} 
\end{equation}
This means we have a collection of identical gauge theories for the fields $\psi_a(b), A_{\mu,a}(b)$, labeled by the scale variable 
$a$, which differ from each other only by the value of the scale-dependent coupling constant $g=g(a)$. It is a matter of choice whether to keep the scale dependence in $g(a)$, or solely in $A_{\mu,a}(b)$. 
The Euclidean action of the multiscale theory takes 
the form
\begin{align}\nonumber 
S_E &=& \frac{1}{C_\chi}\int \frac{dad^db}{a}\Bigl[
\bar\psi_a(b) (\gamma_\mu D_{\mu,a} +\imath m) \psi_a(b) + \\
&+& \frac{1}{4}F_{\mu\nu,a}^A F_{\mu\nu,a}^A 
\Bigr]  
+ \hbox{gauge fixing terms} \label{de2},
\end{align}
where
$$
F_{\mu\nu,a}^A = \partial_\mu A^A_{\nu,a} - \partial_\nu A^A_{\mu,a} - g f^{ABC} A^B_{\mu,a}A^C_{\nu,a}. 
$$
The difference between the standard quantum field  theory formalism 
and the field theory with action \eqref{de2}, defined on the affine group, consists  
in changing the integration measure from $d^dx$ to the left-invariant measure on the affine group [Eq.\eqref{hm}]. So, the 
generating functional can be written in the form 
\begin{equation}
Z[J_{a}(b)] = \int \cD \Phi_a(b) e^{-S_E[\Phi] + \int \frac{dad^4b}{C_\chi a} \Phi_a(b)J_a(b)}, 
\end{equation} 
where $\Phi_a(b) = (A_{a,\mu}(b),\psi_a(b),\ldots)$ is the full set of all scale-dependent fields present in the theory. Since the ''Lagrangian'' in the action \eqref{de2}, for each fixed value of $a$, has exactly the same form as that 
in standard theory, the Faddeev-Popov gauge-fixing procedure \cite{FP1967} can be introduced to the scale-dependent theory in a straightforward way.

\subsection{Feynman diagrams}
The same as in wavelet-regularization of a local theory, described in {\em Sec. \ref{sqft:sec}}, here 
we understand the physically observed fields as the sums of scale components from the observation scale $A$ to infinity \cite{Altaisky2010PRD}:
$$
\psi^A(x) = \frac{1}{C_\chi} \int_A^\infty \frac{da}{a}\int d^db 
\frac{1}{a^d}\chi\left(\frac{x-b}{a} \right) \psi_a(b).
$$
The free-field Green functions at a given scale $a$ are projections of the ordinary Green function to the scale $a$ performed by the $\chi$ wavelet filters:
$$
G_{a_1,\ldots,a_n}(k_1,\ldots,k_n) = \tilde{\bar\chi}(a_1k_1)\ldots
\tilde{\bar\chi}(a_nk_n)G(k_1,\ldots,k_n).$$
The interacting-field Green functions, according to 
the action [Eq.\eqref{de2}], can be constructed if 
we provide the equality of all scale arguments by ascribing the multiplier 
$g(a) \prod_i \delta (\ln a_i - \ln a)$ to each vertex, and 
$ \delta (\ln a_i - \ln a_j)$ to each line of the Feynman diagram. This is different from the  local theory,  
described in {\em Sec. \ref{sqft:sec}}, where all scale components do interact with each other. Now we 
do not yield the cutoff factor $f^2(\cdot)$ on each internal line, with $f(x)$ given by the scale integration
\eqref{cutf1}. Instead, we have to put the wavelet filter modulus squared on each internal line. This suppresses 
not only the UV divergences, {\em but also the IR divergences}. As a result, we arrive at the following diagram technique,
which is (up to the above mentioned cutoff factors), identical to standard Feynman rules for Yang-Mills theory; see, e.g., Ref. \cite{Ramond1989}.

The propagator for the spin-half fermions: 
\begin{equation}\nonumber 
\feynmandiagram [horizontal=c to d] { 
    c[particle=c] -- [fermion,edge label=p] d[particle=d],
  };= \imath \delta_{cd} \frac{\slashed{p}-m}{p^2+m^2} |\tilde{\chi}(ap)|^2, 
\end{equation}
where  $c,d$ are the indices of the fermion representation of the gauge group.

The propagator of the gauge field (taken in the Feynman gauge): 
\begin{equation}\nonumber 
\feynmandiagram [horizontal=A to B] { 
    A[particle=A] -- [boson,momentum=p] B[particle=B],
  };= \delta_{AB}\frac{1}{p^2}\delta_{\mu\nu}  |\tilde{\chi}(ap)|^2 
\end{equation}

The gluon to fermion coupling:
\begin{equation}\nonumber 
\begin{tikzpicture}[baseline=(a)]
\begin{feynman}[horizontal = (a) to (b)]
\vertex (a);
\vertex [above left =1cm of a] (c) {\(c\)};
\vertex [below left =1cm of a] (d) {\(d\)};
\vertex [right = 1cm of a] (b) {\(A\)};
\diagram*{
(d) -- [fermion] (a) -- [fermion] (c),
(a) -- [boson] (b),
};
\end{feynman}
\end{tikzpicture}
= -\imath g(a) \gamma_\mu (T^A)_{cd}
\end{equation}
The three-gluon vertex:
\begin{align}\nonumber 
\begin{tikzpicture}[baseline=(d)]
\begin{feynman}
\vertex (d);
\vertex [above=1cm of d] (b)        {\(B,\nu\)};
\vertex [below left = 1cm of d] (a) {\(A,\mu\)};
\vertex [below right = 1cm of d] (c) {\(C,\rho\)};
\diagram*{
 (b) -- [boson,momentum=q] (d),
 (a) -- [boson,momentum=p] (d),
 (c) -- [boson,momentum=r] (d),
};
\end{feynman}
\end{tikzpicture}
=-\imath g(a) f^{ABC}\bigl[
(r_\mu-q_\mu) \delta_{\nu,\rho} + \\
+ (q_\rho-p_\rho)\delta_{\mu\nu} + (p_\nu-r_\nu) \delta_{\rho\mu} 
\bigr] \label{g3bare}
\end{align}
All momenta are incident to the vertex: $p+q+r=0$.

\begin{widetext}
Similarly, for the four-gluon vertex:
\begin{align*} 
\begin{tikzpicture}[baseline=(e)]
\begin{feynman}[inline=(e)]
\vertex (e);
\vertex [below = 1cm of e] (a) {\(A,\mu\)};
\vertex [above = 1cm of e] (b) {\(B,\nu\)};
\vertex [left  = 1cm of e] (c) {\(C,\rho\)};
\vertex [right = 1cm of e] (d) {\(D,\sigma\)};
\diagram*{
{(a),(b),(c),(d)} -- [boson] (e),
};
\end{feynman}
\end{tikzpicture}
=-g^2(a)\Bigl[ f^{ABE}f^{CDE}(\delta_{\mu\rho}\delta_{\nu\sigma} 
- \delta_{\nu\rho}\delta_{\mu\sigma})  
+ f^{CBE}f^{ADE}(\delta_{\mu\rho}\delta_{\nu\sigma} - \delta_{\nu\mu}\delta_{\rho\sigma}) \\
 + f^{DBE}f^{CAE}(\delta_{\sigma\rho}\delta_{\nu\mu}-\delta_{\nu\rho}\delta_{\mu\sigma})
\Bigr].
\end{align*}
\end{widetext}
The ghost propagator:
\begin{equation}\nonumber
\feynmandiagram [nodes=circle,small, horizontal=A to B] { 
    A -- [scalar,momentum=p] B,
  };= -\imath \frac{\delta^{AB}}{p^2}  |\tilde{\chi}(ap)|^2.
\end{equation}

\begin{widetext}
The gluon-to-ghost interaction vertex:
\begin{align*}\nonumber
\begin{tikzpicture}[baseline=(d)]
\begin{feynman}[inline=(d)]
\vertex (d);
\vertex [below right=1cm of d] (a) {\(A\)};
\vertex [below left =1cm of d] (b) {\(B\)};
\vertex [above   =1cm    of d] (c) {\(C,\mu\)};
\diagram*{
(a) -- [scalar,momentum=p] (d),
(b) -- [scalar,momentum=q] (d),
(c) -- [boson,momentum=r] (d),
};
\end{feynman}
\end{tikzpicture}
=\frac{1}{2}g(a)f^{ABC}(r_\mu+p_\mu-q_\mu)
=-g(a)f^{ABC}q_\mu,
\end{align*}
\end{widetext}
with $r+p+q=0$.

For simplicity, in the following calculations I use the first derivative of Gaussian as a mother  wavelet [Eq.\eqref{g1f:eq}], which 
provides the cancellation of both the UV and the IR divergences by virtue of $|\tilde{\chi}(\cdot)|^2$ on each propagator line. For the chosen wavelet [Eq. 
\eqref{g1f:eq}], the wavelet cutoff factor is  
\begin{equation}
F_a(p) = (ap)^2 e^{-a^2p^2} \label{g1kf}
\end{equation}
for each line of the diagram, calculated for the scale $a$ of the considered model.

\subsection{Scale dependence of the gauge coupling constant}
To study the scale dependence of the gauge coupling constant we can start with a pure gauge field theory without fermions, along the lines of Ref.\cite{GW1973prd}. 
The total one-loop contribution to three gluon interaction is given by the diagram equation \eqref{g3l1:fig}:
\begin{widetext}
\begin{equation}
\begin{tikzpicture}[baseline=(a)]
	\begin{feynman}[inline=(a)]
	\vertex [blob] (a) at (0,0) {\contour{white}{}};
	\vertex[above=1.5cm of a](i3){\(C\)};
	\vertex[below left = 1.5cm of a] (i1) {\(B\)};
	\vertex[below right= 1.5cm of a] (i2) {\(A\)};
		\diagram*{
		{(i3),(i2),(i1)} -- [gluon] (a),
		};
	\end{feynman}
\end{tikzpicture}
=
\begin{tikzpicture}[baseline=(d)]
\begin{feynman}
\vertex (d);
\vertex [above=1cm of d] (b)        {\(C\)};
\vertex [below left = 1cm of d] (a) {\(B\)};
\vertex [below right = 1cm of d] (c) {\(A\)};
\diagram*{
 (b) -- [gluon] (d),
 (a) -- [gluon] (d),
 (c) -- [gluon] (d),
};
\end{feynman}
\end{tikzpicture}
+
\begin{tikzpicture}[baseline=(b1)]
\begin{feynman}[inline=(b1)]
\vertex (a);
\vertex [above=0.8cm of a](i3){\(C\)};
\vertex [below left  = 0.8cm of a]  (b1);
\vertex [below right = 0.8cm of a]  (b2);
\vertex [below left  = 0.8cm of b1] (i1) {\(B\)};
\vertex [below right = 0.8cm of b2] (i2) {\(A\)};
	\diagram*{
		(i3)-- [gluon] (a) -- [gluon] (b1) -- [gluon] (b2) -- [gluon] (a),
		(i1) -- [gluon] (b1), (i2) -- [gluon] (b2),
		};
\end{feynman}
\end{tikzpicture}
+ \frac{1}{2}\Bigl[
\begin{tikzpicture}[baseline=(b)]
\begin{feynman}[inline=(b)]
	\vertex (a) at (0,0);
	\vertex[above=0.8cm of a](i3){\(C\)};
\vertex[below=1cm of a](b);
\vertex[below left = 0.8cm of b] (i1) {\(B\)};
\vertex[below right= 0.8cm of b] (i2) {\(A\)};
\diagram*{
(i3) -- [gluon] (a) -- [gluon,half right] (b),
(i2) [particle=\(A\)] -- [gluon] (b),
(i1)[particle={\(B\)}] -- [gluon] (b) -- [gluon, half right] (a),
};
\end{feynman}
\end{tikzpicture}
+ \hbox{permutations} \Bigr]  
+ \hbox{ghost loops} \label{g3l1:fig}
\end{equation} 
\end{widetext}
In standard QCD, theory the one-loop contribution to the three-gluon vertex is calculated in the Feynman gauge \cite{BC1980}. This was later generalized to an arbitrary covariant gauge \cite{DOT1996}. These  known results, being general in kinematic structure, are based on dimensional regularization, and thus are determined by the divergent parts of integrals. Different corrections to the perturbation expansion based on analyticity have been proposed \cite{SS2007e,BMS2010},
but they are still based on divergent graphs. In this context,  QCD is 
often considered as an {\em effective} theory, which describes 
the low-energy limit for a set of asymptotically observed fields, 
obtained by integrating out all heavy particles \cite{Georgi1993}. The effective theory is believed to be derivable from a future unified theory, which includes gravity.

The essential artifact of renormalized QCD is the logarithmic 
decay of the running coupling constant $\alpha_s(Q^2)$ at infinite 
momentum transfer $Q^2\,\to\,\infty$, known as asymptotic freedom. 
With the help of $\overline{MS}$, the calculations are available up to the five-loop approximation \cite{Baikov2017}.

In the present paper, I do not pretend to derive the logarithmic law.
Instead, I have shown, that if our understanding of gauge invariance is true in an arbitrary functional basis, based on a Lie group representation, we use to measure physical fields, the resulting theory is finite by construction. 
The restriction of calculations to the Feynman gauge and the specific form of the mother  wavelet are technical simplifications, with which we proceed to make the results viewable. 
  
The first term on the rhs of Eq.\eqref{g3l1:fig} is the unrenormalized three-gluon vertex [Eq.\eqref{g3bare}]. The second 
graph is the gluon loop shown in Fig.~\ref{3g3:pic}:
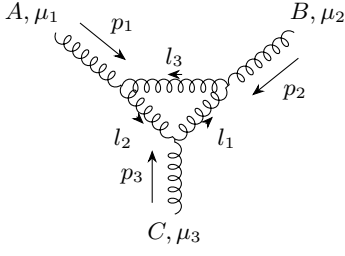
\begin{figure}
\begin{tikzpicture}
\begin{feynman}[vertical= (e) to (f)] 
	\vertex (e);
	\vertex [below=1cm of e] (f) {\(C,\mu_3\)};
	\vertex [above right=1cm of e] (c);
	\vertex [above right=1cm of c] (d) {\(B,\mu_2\)};
	\vertex [above left =1cm of e] (b);
	\vertex [above left =1cm of b] (a) {\(A,\mu_1\)};	
	\diagram*{ 
    (a) -- [gluon,momentum=\(p_1\)] (b), 
    (d) -- [gluon, momentum=\(p_2\)] (c) -- [gluon, rmomentum={[arrow shorten=0.2mm, arrow distance=1.5mm]\(l_3\)}] (b)-- [gluon,rmomentum={[arrow shorten=0.2mm, arrow distance=1.5mm]\(l_2\)}] (e) -- [gluon, rmomentum={[arrow shorten=0.2mm, arrow distance=1.5mm]\(l_1\)}] (c), 
    (f) -- [gluon,momentum=\(p_3\)] e;
    }; 
\end{feynman}
\end{tikzpicture}
\caption{Gluon loop contribution to the three-gluon vertex; $p_1+p_2+p_3=0$.}
\label{3g3:pic}
\end{figure}
Its value is 
\begin{equation}
 \Gamma^{ABC}_{\mu_1\mu_2\mu_3}=-\imath g^3(a) \frac{C_A}{2} f^{ABC} V_{\mu_1,\mu_2,\mu_3}^\mathrm{one-loop}(p_1,p_2,p_3),  
    \label{g3v}
\end{equation}
where the common color factor is $C_A=2T_F N_C$, $N_C$ is the number of colors, and $T_F=\frac{1}{2}$ is the usual normalization of generators in fundamental representation; 
 see, e.g., Ref.\cite{Grozin2007}.

\subsubsection{Gluon loop contribution}
We calculate 
the one-loop tensor structure $V_{\mu_1,\mu_2,\mu_3}^\mathrm{one-loop}(p_1,p_2,p_3)$  
in the Feynman gauge. 
After symmetrization of the loop momenta in diagram \eqref{g3v},
\begin{align*}
l_1 = f + \frac{p_3-p_2}{3}, 
l_2 = f + \frac{p_1-p_3}{3}, 
l_3 = f + \frac{p_2-p_1}{3},
\end{align*} 
the tensor structure of the diagram takes the form 
\begin{align}\nonumber 
V_{\mu_1,\mu_2,\mu_3}^\mathrm{one-loop}(p_1,p_2,p_3,f)=V_{\mu_1,\alpha,\beta}(p_1,l_3,-l_2)\\ 
\times V_{\alpha,\mu_2,\delta}(-l_3,p_2,l_1)V_{\delta,\mu_3,\beta}(-l_1,p_3,l_2),
\end{align}
where 
\begin{align} \nonumber 
V_{\mu_1,\mu_2,\mu_3}(p_1,p_2,p_3):=(p_{3,\mu_1}-p_{2,\mu_1})\delta_{\mu_2,\mu_3} + \\  
+(p_{1,\mu_2}-p_{3,\mu_2})\delta_{\mu_3,\mu_1} +(p_{2,\mu_3}-p_{1,\mu_3})\delta_{\mu_1,\mu_2} \label{ts3}
\end{align}
is the tensor structure of the three-gluon interaction vertex [Eq.\eqref{g3bare}].

The tensor structure of Eq.~\eqref{g3v}
can be represented as a sum of two terms: the first term is free of 
loop momentum $f$, and the second term is quadratic in it:
$$
V_{\mu_1,\mu_2,\mu_3}^\mathrm{one-loop}(p_1,p_2,p_3,f)=V^0(p_1,p_2,p_3) + V^1(p_1,p_2,p_3,f)
$$
with 
\begin{align*}\nonumber 
V^1_{\mu_1,\mu_2,\mu_3}(p_1,p_2,p_3,f) = 3\bigl[
f_{\mu_1}f_{\mu_3} (p_{1,\mu_2}-p_{3,\mu_2}) + \\ \nonumber 
+ f_{\mu_1}f_{\mu_2} (p_{2,\mu_3}-p_{1,\mu_3}) + f_{\mu_2}f_{\mu_3} (p_{3,\mu_1}-p_{2,\mu_1}) 
\bigr] + \\ \nonumber 
+ \frac{7}{3}f^2 \bigl[
(p_{3,\mu_1}-p_{2,\mu_1})\delta_{\mu_2,\mu_3}   
+(p_{1,\mu_2}-p_{3,\mu_2})\delta_{\mu_3,\mu_1} +\\ \nonumber  
+(p_{2,\mu_3}-p_{1,\mu_3})\delta_{\mu_1,\mu_2}
\bigr] + \frac{2}{3} \bigl[
\delta_{\mu_1\mu_2} f_{\mu_3} f_\alpha (p_{2,\alpha}-p_{1,\alpha}) +  \\
\delta_{\mu_1\mu_3} f_{\mu_2} f_\alpha (p_{1,\alpha}-p_{3,\alpha}) + \delta_{\mu_2\mu_3} f_{\mu_1} f_\alpha (p_{3,\alpha}-p_{2,\alpha}) 
\bigr]
\end{align*}
Integrating the equation $V^1_{\mu_1,\mu_2,\mu_3}(p_1,p_2,p_3,f)$ with the Gaussian weight  we substitute $f_\mu f_\nu \to \frac{\delta_{\mu\nu}}{d}f^2$ into the Gaussian integral $\int e^{-\zeta f^2} f^2 d^df = \frac{d}{2}\pi^\frac{d}{2} \zeta^{-\frac{d}{2}-1}$. With $\zeta=3a^2,d=4$, this gives the tensor structure 
\begin{align*}\nonumber 
V^1(p_1,p_2,p_3) = \frac{13}{864\pi^2} 
e^{-\frac{2}{9}a^2[p_1^2+p_2^2 + p_3^2 - p_1p_2 - p_1 p_3 - p_2 p_3]}\\
\times V_{\mu_1,\mu_2,\mu_3}(p_1,p_2,p_3).
\end{align*}
The part of the tensor structure that does not contain $f$ contributes a term proportional to the Gaussian integral 
$\int e^{-\zeta f^2} d^df = \left(\frac{\pi}{\zeta} \right)^\frac{d}{2}$. This gives 
$$
\frac{a^2}{144\pi^2}e^{-\frac{2}{3}a^2[p_1^2+p_2^2+p_1p_2]}V^0_{\mu_1,\mu_2,\mu_3}(p_1,p_2,p_3=-p_1-p_2)
$$
where 
\begin{align*}
V^0_{\mu_1,\mu_2,\mu_3}(p_1,p_2)= \frac{4}{3}(p_{2,\mu_1} p_{2,\mu_2}p_{2,\mu_3} - p_{1,\mu_1} p_{1,\mu_2}p_{1,\mu_3}) \\
+ \frac{5}{3}(p_{2,\mu_1} p_{2,\mu_2}p_{1,\mu_3} - p_{1,\mu_1} p_{1,\mu_2} p_{2,\mu_3}) \\
+ \frac{2}{3}(p_{2,\mu_2} p_{2,\mu_3}p_{1,\mu_1} - p_{1,\mu_1} p_{1,\mu_3} p_{2,\mu_2}) \\
+ \frac{1}{3}(p_{1,\mu_2} p_{1,\mu_3}p_{2,\mu_1} - p_{2,\mu_1} p_{2,\mu_3} p_{1,\mu_2}) \\
+ \frac{37}{27} \delta_{\mu_1\mu_2}p_1p_2 (p_{2,\mu_3}-p_{1,\mu_3})+ \frac{58}{27}\delta_{\mu_1\mu_2}
(p_1^2 p_{2,\mu_3} - p_2^2 p_{1,\mu_3}) \\
+ \frac{5}{27}(\delta_{\mu_2\mu_3}p_1^2 p_{1,\mu_1}) - \delta_{\mu_1\mu_3}p_2^2 p_{2,\mu_2}- \delta_{\mu_1\mu_2}p_2^2 p_{2,\mu_3} )\\
+ \frac{32}{27} (
\delta_{\mu_1\mu_3}p_{1,\mu_2}(p_1^2+p_1p_2)-\delta_{\mu_2\mu_3}p_{2,\mu_1}(p_2^2+p_1p_2)
) \\
+\frac{16}{27}(
\delta_{\mu_1\mu_3}p_1^2 p_{2,\mu_2} - \delta_{\mu_2\mu_3}p_2^2 p_{1,\mu_1} 
) \\
+\frac{53}{27}(
\delta_{\mu_1\mu_3}p_2^2 p_{1,\mu_2} - \delta_{\mu_2\mu_3}p_1^2 p_{2,\mu_1} 
)\\
+\frac{47}{27}p_1 p_2(
\delta_{\mu_2\mu_3}p_{1,\mu_1} - \delta_{\mu_1\mu_3}p_{2,\mu_2}
)
\end{align*}
Summing these two terms,  we get
\begin{align} \nonumber 
\Gamma_{\mu_1\mu_2\mu_3}^{ABC}(p_1,p_2) = -\imath g^3(a) \frac{C_A}{2} \frac{f^{ABC}}{144\pi^2} \times \\
\times e^{-\frac{2}{3}a^2 (p_1^2+p_2^2+p_1p_2)}\Bigl[ a^2V^0_{\mu_1\mu_2\mu_3}(p_1,p_2) + \\
\nonumber + \frac{13}{6}V_{\mu_1\mu_2\mu_3}(p_1,p_2,-p_1-p_2)\Bigr],
\end{align}
where $V_{\mu_1\mu_2\mu_3}$, given by Eq.\eqref{ts3}, is the tensor 
structure of the unrenormalized three-gluon vertex.

\subsubsection{Contribution of four-gluon vertex}
\begin{figure}
\begin{tikzpicture}
\begin{feynman}[vertical= (a) to (b)] 
\vertex (a);
\vertex [above=1cm of a] (i3) {\(C,\mu_3\)};
\vertex [below=1.5cm of a] (b);
\vertex [below left  = 1cm of b] (i1) {\(B,\mu_2\)};
\vertex [below right = 1cm of b] (i2) {\(A,\mu_1\)};
\diagram*{ 
    (i3) -- [gluon,momentum=\(p_3\)] (a) -- [gluon, half right, momentum=\(f\)] (b), 
    (i2)-- [gluon, momentum=\(p_1\)] (b),
    (i1)-- [gluon, momentum=\(p_2\)] (b) -- [gluon, half right, edge label=\(D\)] (a),
    };
\end{feynman}
\end{tikzpicture}
\caption{One-loop contribution to the three-gluon vertex provided by four-gluon interaction.}
\label{3g4:pic}
\end{figure}
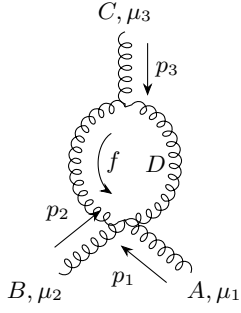

The next one-loop contribution to the three-gluon vertex comes from the diagrams with four-gluon interaction, of the type shown in Fig.~\ref{3g4:pic}.
In the case of four-gluon contribution the common color factor cannot be factorized: instead there are three similar diagrams with gluon loops inserted in each gluon leg: $p_3$, $p_2$ and $p_1$, respectively. The case of $p_3$ is shawn in Fig.~\ref{3g4:pic}. 

The one-loop contribution  to a three-gluon vertex shown in Fig.~\ref{3g4:pic} can be easily calculated taking into account that the squared momenta 
in gluon propagators are cancelled by wavelet factors [Eq.\eqref{g1kf}]. This gives 
\begin{align*}
a^4 \int (-\imath g(a)) f^{DEC} \bigl[ (2p_3-f)_\delta \delta_{\mu_3\epsilon} 
+ (-f-p_3)_\epsilon \delta_{\mu_3\delta}+\\ 
+ (2f-p_3)_{\mu_3} \delta_{\epsilon\delta}\bigr]\times \\
\times (-g(a))^2 \bigl[ f^{AEX}f^{BDX}(\delta_{\mu_1 \mu_2}\delta_{\epsilon\delta}-\delta_{\epsilon\mu_2}\delta_{\mu_1\delta})+\\
+ f^{BEX}f^{ADX}(\delta_{\mu_1\mu_2}\delta_{\epsilon\delta}-\delta_{\epsilon\mu_1}\delta_{\mu_2\delta}) + \\
+ f^{DEX}f^{BAX} (\delta_{\delta\mu_2} \delta_{\epsilon\mu_1}-\delta_{\epsilon\mu_2}\delta_{\mu_1\delta})\bigr] \times \\
\times \exp\bigl(-a^2f^2-a^2(f+p_1+p_2)^2 \bigr) \frac{d^4f}{(2\pi)^4}.
\end{align*} 
The presence of four-gluon interaction does not allow for the factorization of the common color factor. Instead, there are three different terms in color 
space:
\begin{align}\nonumber 
f^{DEC}f^{AEX}f^{BDX}=-\frac{C_A}{2}f^{ABC}, 
\\
f^{DEC}f^{BEX}f^{ADX}=+\frac{C_A}{2}f^{ABC}, \label{T2c}\\
\nonumber 
f^{DEC}f^{DEX}f^{BAX}=-C_Af^{ABC} 
\end{align}
with the normalization condition 
\begin{align*}
f^{ACD}f^{BCD}=C_A \delta_{AB}.
\end{align*}

There are two Gaussian integrals contributing to the diagram 
shown in Fig.~\ref{3g4:pic}:
\begin{align*}
I(s)    &=& \int \frac{d^4f}{(2\pi)^4} e^{-2a^2f^2 -2a^2sf} = \frac{1}{64\pi^2a^4} e^\frac{a^2s^2}{2}, \\
I_\mu(s)&=& \int \frac{d^4f}{(2\pi)^4} f_\mu e^{-2a^2f^2 -2a^2sf} = -\frac{s_\mu}{128\pi^2a^4} e^\frac{a^2s^2}{2},
\end{align*}
where  $s=p_1+p_2$. 

Thus we can express the tensor coefficients at the three terms in 
Eq.\eqref{T2c} as 
\begin{align}\nonumber 
T_1 &=& -\frac{3}{2}\delta_{\mu_1\mu_3} s_{\mu_2} + \frac{3}{2}\delta_{\mu_2\mu_3} s_{\mu_1}, \\
T_2 &=& +\frac{3}{2}\delta_{\mu_1\mu_3} s_{\mu_2} - \frac{3}{2}\delta_{\mu_2\mu_3} s_{\mu_1}, \\  \nonumber
T_3 &=& 3 (\delta_{\mu_2\mu_3} s_{\mu_1} -\delta_{\mu_1\mu_3}s_{\mu_2}),
\end{align}
respectively.
The sum of all three terms $-\frac{C_A}{2}f^{ABC}T_1 + \frac{C_A}{2}f^{ABC}T_2 -C_A f^{ABC} T_3$ gives 
$$ \frac{9}{2} C_A f^{ABC}[ \delta_{\mu_1\mu_3}s_2 - \delta_{\mu_2\mu_3}s_{\mu_1}], $$
and thus the whole integral 
\begin{equation}
V_{\mu_1\mu_2\mu_3}^{ABC}(s) = \frac{\imath g^3(a)}{64\pi^2} e^\frac{-a^2s^2}{2} \frac{9C_A}{2} f^{ABC}[ \delta_{\mu_1\mu_3}s_{\mu_2} - \delta_{\mu_2\mu_3}s_{\mu_1}].
\end{equation}
Two more contributing diagrams, symmetric to Fig.~\ref{3g4:pic}, are different from the above calculated 
$V(A,\mu_1,p_1;B,\mu_2,p_2;C,\mu_3,p_3)$ only by changing $B,\mu_2,p_2 \leftrightarrow C,\mu_3,p_3$ and 
$ A,\mu_1,p_1 \leftrightarrow C,\mu_3,p_3$, respectively. This gives two more terms 
\begin{align*}
V_{\mu_1\mu_2\mu_3}^{ABC}(t) = \frac{\imath g^3(a)}{64\pi^2} e^\frac{-a^2t^2}{2} \frac{9C_A}{2} f^{ACB}[ \delta_{\mu_1\mu_2}t_{\mu_3} - \delta_{\mu_2\mu_3}t_{\mu_1}], \\
V_{\mu_1\mu_2\mu_3}^{ABC}(u) = \frac{\imath g^3(a)}{64\pi^2} e^\frac{-a^2u^2}{2} \frac{9C_A}{2} f^{CBA}[ \delta_{\mu_1\mu_3}u_{\mu_2} - \delta_{\mu_2\mu_1}u_{\mu_3}],
\end{align*}
where $t=p_1+p_3=-p_2$ and $u=p_2+p_3=-p_1$. 
Taking into account the common topological factor $\frac{1}{2}$ standing before all  these diagrams in \eqref{g3l1:fig}, finally we get 
\begin{align}\nonumber 
\Gamma^{ABC}_{\mu_1\mu_2\mu_3}(p_1,p_2) = \imath \frac{g^3(a)}{256\pi^2} 9 C_A f^{ABC} \Bigl[ 
e^\frac{-a^2s^2}{2} (\delta_{\mu_1\mu_3}s_{\mu_2} \\  - \delta_{\mu_2\mu_3}s_{\mu_1}) + 
e^\frac{-a^2p^2_2}{2} (\delta_{\mu_1\mu_2}p_{2,\mu_3} - \delta_{\mu_2\mu_3}p_{2,\mu_1}) \\ 
\nonumber + 
e^\frac{-a^2p^2_1}{2} (\delta_{\mu_1\mu_3}p_{1,\mu_2} - \delta_{\mu_1\mu_2}p_{1,\mu_3}) \Bigr],
\end{align}
where  $t=-p_2,u=-p_1,p_3=-p_1-p_2$.

\subsubsection{Ghost loop contribution}
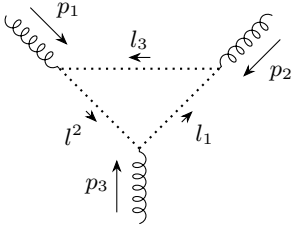
\begin{figure}
\begin{tikzpicture}
\begin{feynman}[vertical= (e) to (f)]
\vertex (e);
\vertex [below=1cm of e] (f);
\vertex [above left = 1.5cm of e] (b);
\vertex [above left = 1cm of b] (a);
\vertex [above right= 1.5cm of e] (c);
\vertex [above right= 1 cm of c] (d);
\diagram*{ 
    (a) -- [gluon,momentum=\(p_1\)] (b), 
    (d) -- [gluon, momentum=\(p_2\)] (c) -- [ghost, rmomentum={[arrow shorten=0.2mm, arrow distance=1.5mm]\(l_3\)}] (b) -- [ghost,rmomentum={[arrow shorten=0.2mm, arrow distance=1.5mm]\(l^2\)}] (e) -- [ghost, rmomentum={[arrow shorten=0.2mm, arrow distance=1.5mm]\(l_1\)}] (c), (f) -- [gluon,momentum=\(p_3\)] (e),
    };
\end{feynman} 
\end{tikzpicture}
\caption{Ghost loop contribution to three-gluon vertex; $p_1+p_2+p_3=0$.}
\label{3gh3:pic}
\end{figure}

The last one-loop contribution not shown in Eq.\eqref{g3l1:fig}, is the ghost loop diagram Fig.~\ref{3gh3:pic}, and one more 
diagram symmetric to it. 
The color factor of the diagram Fig.~\ref{3gh3:pic} is $f^{DEA}f^{FDB}f^{EFC}=-\frac{C_A}{2}f^{ABC}$. The tensor structure for diagram Fig.~\ref{3gh3:pic} is 
$l_{2,\mu_1}l_{3,\mu_2}l_{1,\mu_3}$, and it is $l_{3,\mu_1}l_{1,\mu_2}l_{2,\mu_3}$  for the symmetric diagram \cite{Grozin2007}. Ghost propagators multiplied by wavelet factors give 
$(-\imath)^3 a^6 e^{-3a^2f^2 - \frac{2}{3}a^2(p_1^2+p_2^2+p_1p_2)}$, and one more $(-1)$ accounts for the fermion loop. 
Finally, this gives 
\begin{align} \nonumber 
\Gamma^{ghost} &=& -\imath g^3(a)\frac{C_A}{2} f^{ABC}
\frac{e^{-\frac{2}{3}a^2 (p_1^2+p_2^2+p_1p_2)}}
{144\pi^2} \times \\
&\times& \bigl[ a^2 V_0 + \frac{1}{18}V_{\mu_1\mu_2\mu_3}(p_1,p_2,p_3=-p_1-p_2)\bigr], \\  \nonumber 
V_0 &=& \frac{1}{27}(p_{1,\mu_3}-p_{2,\mu_3})(p_{1,\mu_2}p_{2,\mu_1}-2p_{1,\mu_1}p_{2,\mu_2})  \\ 
\nonumber &+& 
\frac{4}{27}(
 p_{2,\mu_1}p_{2,\mu_2}p_{2,\mu_3}
-p_{1,\mu_1}p_{1,\mu_2}p_{1,\mu_3}
)  \\
\nonumber &+&
\frac{5}{27} (
  p_{2,\mu_1} p_{2,\mu_2} p_{1,\mu_3} 
- p_{1,\mu_1}p_{1,\mu_2}p_{2,\mu_3}
).
\end{align}

\subsection{Study of simplified 3-gluon vertex $(p,-p,0)$}
To study the scale dependence of the coupling constant let us start with a trivial situation $p_1=p,p_2=-p,p_3=0$. 
The unrenormalized vertex takes the form 
$$
\Gamma_{\mu_1\mu_2\mu_3}^{ABC}(p) = -\imath g(a) f^{ABC} V(p,-p,0), $$ where $$
V(p,-p,0) \equiv  p_{\mu_1}\delta_{\mu_2\mu_3} + p_{\mu_2}\delta_{\mu_1\mu_3}
-2 p_{\mu_3} \delta_{\mu_1\mu_2} .
$$
The triangle gluon loop contribution, shown in Fig.~\ref{3g3:pic}, is 
\begin{align}\nonumber
\Gamma_{\mu_1\mu_2\mu_3}^{ABC,3}(p) &=& -\imath g(a)^3 \frac{C_A}{2} f^{ABC} \frac{e^{-\frac{2}{3}a^2p^2}}{144\pi^2} \times \\ 
&\times& \bigl[a^2V_0 + \frac{13}{6}V(p,-p,0)\bigr] \label{g3pp}
\\   \nonumber 
V_0 &=& \frac{4}{3} p_{\mu_1}p_{\mu_2}p_{\mu_3} - 
\frac{p^2}{27}\bigl( 
5 \delta_{\mu_2\mu_3} p_{\mu_1} + \\   \nonumber
 &+& 5 \delta_{\mu_1\mu_3}p_{\mu_2} + 32 \delta_{\mu_1\mu_2} p_{\mu_3} \bigr).
\end{align}
The contributions containing four-gluon vertexes (without fermions) give 
\begin{align}
\Gamma_{\mu_1\mu_2\mu_3}^{ABC,4}(p) = -\imath \frac{g^3(a)}{256\pi^2} 9C_A f^{ABC} e^{-\frac{a^2p^2}{2}}V(p,-p,0).
\end{align}
The contributions of two ghost loops give 
\begin{align}\nonumber 
\Gamma^{ghost}_{\mu_1\mu_2\mu_3}(p,-p,0) = -\imath g^3(a)\frac{C_A}{2} \frac{f^{ABC}}{144\pi^2} e^{-\frac{2}{3}a^2p^2} \times \\
\times \frac{1}{9} \bigl[ a^2 \frac{4}{3} p_{\mu_1}p_{\mu_2}p_{\mu_3}
 + \frac{1}{2}V(p,-p,0) \bigr]. \label{g3gh}
\end{align} 


Therefore, due to the use of a localized wavelet basis with a window width of size $a$, we obtain an exponential decay of the vertex function proportional to $p^2$. 

The gauge interaction in the action functional [Eq.\eqref{de2}] is not identical to that of local gauge theory \eqref{de1}.
At this point I cannot definitely claim that physical observables are integrals of the form $\int_A^\infty \frac{da}{a} F[\phi_a(b)]$. 
If the parameter $A$ of a wavelet-regularized local theory 
\eqref{gfw} were a counterpart of a $1/\mu$ normalization scale, 
in our theory with scale-dependent gauge invariance the 
scale parameter $a$ should be treated as an independent coordinate on a $(d+1)$-dimensional group manifold ($a,\mathbf{x}$), with 
the scale transformations given by the generator $D=a \partial_a$. 

Using the simplified vertex contributions [Eqs.\eqref{g3pp}--\eqref{g3gh}] of the one-loop scale-dependent Yang-Mills theory we can estimate the renormalization of the coupling constant $g$ in the considered theory with scale-dependent 
gauge invariance [Eq.\eqref{grot}]. Since the scale $a$ in such a theory plays the role of the normalization scale $1/\mu$ of common 
models, we can calculate the $\beta$ function 
$$
\beta = - a^2 \left. \frac{\partial g}{\partial a^2}\right|_{g_0=const}
$$
from the equality $g_0 = Z_1 g$, with 
\begin{equation}
Z_1 = 1 + \frac{g^2 C_A}{16\pi^2} \left[
\frac{10}{81} e^{-\frac{2}{3}a^2p^2} + \frac{9}{16} e^{-\frac{1}{2}a^2p^2}
\right]
\end{equation}
calculated from the one-loop expansion \eqref{g3l1:fig} with the  vertex contributions [Eqs.\eqref{g3pp}--\eqref{g3gh}].
This gives 
\begin{equation}
\beta = - g a^2 \frac{\partial}{\partial a^2} \frac{1}{Z_1}.
\label{beta}
\end{equation}

The equation \eqref{beta} differs from standard renormalization schemes by the absence of the  
factor $Z_3^\frac{3}{2}$ for field renormalization. This is because each of the scale-dependent fields $A_{\mu,a}(b)$ dwells on its own scale $a$, and is not subjected to renormalization \cite{Altaisky2016PRD}. Taking the scale derivative in Eq.\eqref{beta} explicitly we get 
\begin{equation}
\beta = - \frac{g^3 C_A (ap)^2}{16\pi^2}
\left[
\frac{20}{243} e^{-\frac{2}{3}a^2p^2} + \frac{9}{32} e^{-\frac{1}{2}a^2p^2}
\right] < 0. \label{beta2}
\end{equation} 
The dependence of this function on dimensionless scale $x=ap$ is shown in Fig.~\ref{beta:pic} below.
\begin{figure}[ht]
\centering \includegraphics[width=7cm]{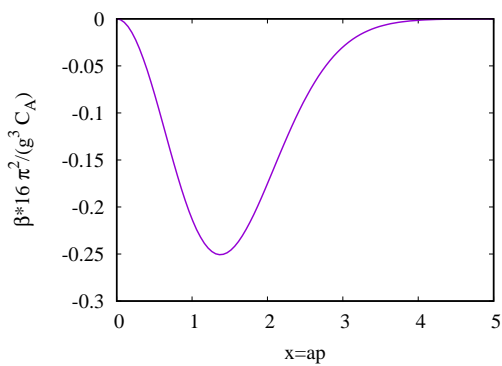}
\caption{Dependence of the beta function [Eq.\eqref{beta2}] on the dimensionless scale $x=ap$. The dependence is shown in units of $\frac{g^3 C_A}{16\pi^2}$.}
\label{beta:pic}
\end{figure}

The factors similar to field renormalization may be required depending on the type of observation -- if we assume the observable quantity to be dependent on $\frac{\langle AAA\rangle}{\langle A\rangle^3}$, where the averaging $\langle \cdot \rangle$ involves  integration over a certain range of scales. 
A more detailed study of the subject is planned for future research.
Since the action In Eq.\eqref{de2} comprises the fields of different scales, which do not interact to each other, to 
derive a phenomenological interpretation of the proposed model we need to study it within a wider framework of the 
Standard Model with the $SU(2) \times U(1) \times SU_c(3)$ gauge group to calculate the observable quantities. 

\section{Conclusion \label{conc:sec}}
The basis of Fourier harmonics, an omnipresent tool of quantum field theory, is just a particular case of the decomposition of the 
observed field $\phi$ with respect to representations $\Omega(g)$ of the symmetry group $G$ responsible for observations. It is commonly assumed that the symmetry group of measurement is a translation group (or, more generally, the Poincar\'e group) the 
representations of which are used. We can imagine, however, that 
the measurement process itself is more complex, and may have symmetries more complex than the Abelian group of translations. 
The simplest generalization is the affine group $G: x'=ax+b$, considered in this paper -- a tool for studying scaling properties of physical systems. 
In this paper,  I have considered the 
possibility to extend quantum field theory models of gauge fields, usually defined in $\R^d$ or Minkowski space, 
to more general space -- the group of affine transformations, which includes not only translations and rotations, 
but also scale transformation. The peculiarity of the scale parameter ($a$) is that the scale transformation generator $D= a \partial_a$, in contrast to coordinate or momentum operators, is not a Hermitian operator. Hence, 
the scale $a$ is not a physical {\em observable} -- it is a parameter of measurement -- say, a scale we use in our measurements. 

Following the previous papers \cite{Altaisky2010PRD,AK2013,Altaisky2016PRD}, introducing explicitly a basis 
$\chi(a,\cdot)$ to describe quantum fields, the current paper presents a gauge theory with a gauge transformation 
defined separately on each scale, $\psi_a(x) \to e^{\imath \Omega_a(x)} \psi_a(x)$. The transformation from the usual 
local fields $\psi(x)$ to the scale-dependent fields, which may be referred to as the scale components of the 
field $\psi$ with respect to the basic function $\chi$ at a given scale $a$, is performed by means of continuous wavelet transform -- a versatile tool of group representation theory. This representation is physically similar to 
coherent state representation \cite{DGM1986}. The Green functions in the scale-dependent theory become finite, for 
both the UV and the IR divergences are suppressed by the wavelet factor   $|\tilde{\chi}(ap)|^2$  on each internal line of the Feynman diagrams.

As a practical example of calculations, the paper presents one-loop correction to the three-gluon vertex in a pure Yang-Mills theory. The calculations are done with the mother wavelet $\chi$ being the first derivative of the Gaussian. The Green functions vanish at high momenta, which is usual for the theories with asymptotic freedom.

 The existence of such a theory is merely an exciting mathematical possibility. The author does not know, which type of interaction takes place in real processes: standard {\em local} gauge theory, where all scales talk to each other due 
to locally defined gauge invariance, or the {\em same-scale} interaction proposed in this paper. This subject needs further investigation -- at least, it seems not less elegant than the existing finite-length and noncommutative geometry models \cite{Freidel2006,Blaschke2010}.  

\section*{Acknowledgement}
The author is thankful  to Drs. A.V.Bednyakov, S.V.Mikhailov and O.V.Tarasov for useful discussions, and to anonymous referee for useful comments.
%

\end{document}